%Paper: hep-th/9206061
%From: IKEMORI%JPNYITP.BITNET@pucc.princeton.edu
%Date: Tue, 16 Jun 92 14:34:56 JST

%%%%%%%%%%%%%%%%%%%%%%%%%%%%%%%%%%%%%%%%%%%%%%%%%%%%%%%%%%%%%%%%%%%%%%%%
%   Extended form method of antifield-BRST formalism
%    for topological quantum field theories,
%%%%%%%%%%%%%%%%%%%%%%%%%%%%%%%%%%%%%%%%%%%%%%%%%%%%%%%%%%%%%%%%%%%%%%%%
\input phyzzx

\date={June, 1992}

\titlepage
\title{\seventeenrm
       Extended form method of antifield-BRST formalism
       for topological quantum field theories
      }

\author{Hitoshi IKEMORI\foot{E-mail address: ikemori@jpnyitp.bitnet}}
\address{Faculty of Economics, Shiga University
 \nextline
1-1-1 Bamba, Hikone, Shiga 522, Japan}

\centerline{and}

\address{Department of Physics,
 Nagoya University\foot{\rm Current address} \nextline
 Nagoya 464-01, Japan}

\abstract
{
 We present a concise method to construct a BRST invariant action
for the topological quantum field theories
in the Batalin-Vilkovisky antifield formalism.
The BV action that is a solution for the master equation is
directly obtained  by means of the extended
forms that involve all the required  ghosts and antifields.
The BV actions for the non-abelian $BF$ theories
(in 4 and higher dimensions) and the Chern-Simons theory are
constructed by means of the extended form method.
An extension of the $BF$ theory in 4-dimensions to include a
``cosmological term'' is also examined and the close connection
with the topological Yang-Mills theory is indicated.
 }

\endpage
%%%%%%%%%%%%%%%%%%%%%%%%%%%%%%%%%%%%%%%%%%%%%%%%%%%%%%%%%%%%%%%%%%%%%%%
\def\tilde{\widetilde}

\noindent{\bf 1. Introduction}

 It is well known that the BRST algebra plays an
important role not only in ordinary gauge theories but also in
 topological quantum field theories (TQFT) [1-4].
The only observables in TQFT are topological invariants
characterized by the BRST cohomologies,
and it is a primary task to construct
the BRST transformations and the BRST invariant actions
in these theories.
There is a useful algorism  presented by Batalin-Vilkovisky (BV)
to construct a BRST invariant gauge fixed action
in a covariant manner [5].
Although the BV antifield-antibracket formalism is
a well constructed algorism,
it is not an easy task to solve the BV master equation
in a direct way especially for highly reducible theories.
It is known, however, that algebraic methods can save one's labors
to obtain the BRST algebra for some kinds of TQFT.
The method of extended differential calculus on
the universal bundle [6-7]
serves as a basis in the case of topological Yang-Mills (TYM) theory
and is applied to the case of $BF$ theories [11].

 In this paper, we present a concise method to construct
a BRST invariant BV action for TQFT based on an
advanced idea along these lines [12].
The BV action that is a solution for the master equation
is directly obtained
by means of the extended differential
forms that involves all the required  ghosts and antifields.
Though the BRST algebra is also derived from a simple condition,
the BV action itself can be obtained without referring
to a concrete form of the algebra.

 The plan of this paper is as follows.
The next section, \S2, deals with an antifield-BRST construction of
non-abelian $ BF $ theory in 4-dimensions
by means of our extended form method.
\S3 makes an indication of connection between
our method and the odd time canonical formulation
of the BV formalism.
In \S4, the BV action for $BF$ theories in arbitrary dimensions
is constructed.
\S5 treats an application to the Chern-Simons theory.
In \S6, an extension of the $BF$ theory in 4-dimensions to include a
``cosmological term'' is examined
and its close connection with a topological Yang-Mills theory
is indicated. The final section, \S7, is for a conclusion.

%\vskip20pt
\endpage

\noindent{\bf 2. $BF$ theory in 4-dimensions}

 Let us begin with the BV-BRST construction for the 4-dimensional
non-abelian $BF$ system
as a simple but non-trivial example suitable for our method.
The $BF$ system [9,10]
is considered as a sort of TQFT which bears a resemblance to
the Chern-Simons theory, whose classical action is
$$
  S[A,B] = \int _{M_4}{\rm Tr} ( B\wedge F )
 \eqno(2.1)
$$
in the differential forms.
The fundamental fields in the action above are Lie algebra valued
1-form $A$ and 2-form $B$,
(where $\ F= dA + {1\over 2}[A,A]\ $ is
the curvature 2-form of connection $A$),
therefore the field equations derived from the action are
$$
  F=0\ ,\quad DB=0\ ,
 \eqno(2.2{\rm a,b})
$$
which requires flat connections as the Chern-Simons theory.

 The action (2.1) is invariant under the transformation
$$
  \delta _\varepsilon A = 0\ ,\quad \delta _\varepsilon B = D\varepsilon
 \eqno(2.3)
$$
as well as the ordinary Yang-Mills gauge transformation
$$
  \delta _\omega A = D\omega \ ,\quad \delta _\omega B = [\omega ,B]\ ,
 \eqno(2.4)
$$
where $\varepsilon $ and $\omega $ are 1-form and 0-form respectively.
The former of these symmetries is on-shell reducible,
which means that it is reduced to $\ \delta _{\varepsilon }B
 = [F,\lambda ] \ $ for
$\ \varepsilon = D\lambda \ $ and vanishes on-shell.

 The first stage reducibility of the $\varepsilon $-symmetry
requires us to introduce not only ghost
but also ghost-for-ghost
(which we will denote as $\psi $ and $\phi $ respectively)
when applying the BV algorism.
It is a matter of course that the $\omega $-symmetry
requires a ghost $c$ as usual.

We will follow the convention
that a $p$-form with ghost number $q$ is called $(p,q)$-form,
therefore we will often attach a
subscript to the form indicating these degrees as $\ \Phi _{(p,q)}\ $
when necessary.
For example, we denote the sequence of ghosts descended from
$\ B=B_{(2,0)}\ $
as $\ \psi =\psi _{(1,1)}\ $ and $\ \phi _{(0,2)}\ $.

 An algebraic method to obtain a BRST transformations for this system
has been presented in ref. [11], that is a sort of extension of
the algebraic derivation in topological Yang-Mills theory.
We will present a new derivation which is a modified and extended
version of this algebraic method.
The basic idea of these methods is an extension of the
exterior derivative to a sum of the usual exterior derivative $\ d\ $
and the BRST operator $\ s\ $,
$$
\tilde d = d + s \ .
 \eqno(2.5)
$$
According to this extension,
the differential form of type $(p,q)$ should be thought
as an extended form of total degree $ p+q $.

 The first attempt in ref. [11] is to define the extended forms
 $\tilde A$ and $\tilde B$ as
$$
 \tilde A = A_{(1,0)}\ ,
\quad \tilde B = B_{(2,0)} + \psi _{(1,1)} +\phi _{(0,2)}\ ,
 \eqno(2.6{\rm a,b})
$$
on counting the degrees of the ghosts descended from $B$.
However, the first candidate for BRST algebra obtained from
the horizontality conditions
$$
  \tilde F = F\ ,\quad \tilde D\tilde B=DB
 \eqno(2.7{\rm a,b})
$$
fails to be nilpotent. Therefore, an improved method is presented
in ref. [11], which is to include the BV antifields
into the extended 1-form $\tilde A$ as
$$
  \tilde A = A_{(1,0)} + B^*_{(2,-1)}
 + \psi ^*_{(3,-2)} +\phi ^*_{(4,-3)}
 \eqno(2.8)
$$
instead of (2.6a), where $B^*$, $\ \psi ^*\ $ and $\ \phi ^*\ $
are the BV antifields for $B$, $\psi $ and $\phi $ respectively.
It should be noticed that
the BV antifield for $(p,q)$-form
in $D$-dimensional spacetime are defined as a $(D-p,-1-q)$-form:
$$
   \Phi _{(p,q)} \ \rightarrow \ \Phi ^*_{(D-p,-1-q)}\ .
 \eqno(2.9)
$$
This time, conditions
$$
  \tilde F=0\ ,\quad \left( \tilde D\tilde B\right) _{q\geq 1}=0
 \eqno(2.10{\rm a,b})
$$
proposed by them lead to nilpotent BRST transformations and
they construct a BRST invariant BV action so as to derive
the transformations by this BV action considering it as a generator.
What we call BV action is a solution for the master equation
which contains antifields and
is still to be gauge fixed by introducing a gauge fermion together with
antighosts and multipliers.

 Although their method seems to work well,
there is a scope for improvement to settle some unsatisfactory aspects.
We note that there is a restriction on ghost number
(${q\geq 1}$) in the expansion of the
condition (2.10b) to produce BRST transformations
and that the BV action is only determined by referring them
because of lack of direct construction using the extended forms.

 We present a more improved method to settle these issues by means of
full extension of the forms.
We define extended forms $\tilde A$ and $\tilde B$ as
$$
 \eqalignno
 {
 \tilde A &= c_{(0,1)} + A_{(1,0)}
         + B^*_{(2,-1)} + \psi ^*_{(3,-2)} +\phi ^*_{(4,-3)}\ ,
     &(2.11{\rm a}) \cr
 \tilde B &= c^*_{(4,-2)} + A^*_{(3,-1)} + B_{(2,0)}
                              + \psi _{(1,1)} +\phi _{(0,2)}\ ,
     &(2.11{\rm b}) \cr
 }
$$
where $c$ is the ordinary ghost for Yang-Mills gauge and
the anti-fields $A^*$ and $c^*$ for $A$ and $c$ are also included.
It should be remarked that this is a full extension in 4 dimensions
and that $\tilde A$ and $\tilde B$ can be regarded as antifields
of each other, which is a key observation in our construction.

Our proposal for conditions to derive the BRST transformations are
simply
$$
  \tilde F=0\ ,\quad \tilde D\tilde B=0
 \eqno(2.12{\rm a,b})
$$
without any restriction on the expansion in ghost number.
These conditions take the same form as
the field equations (2.2a,b) from the classical action.
The expansion in ghost number of these conditions (2.12a,b) produces
$$
  \tilde F=0 \quad \Rightarrow \
\left\{ \eqalign
 {
    sc   &=  - {1\over 2}[c,c]                        \cr
    sA   &=  - Dc                                 \cr
    sB^* &=  - F - [c,B^*]                        \cr
    s\psi ^* &= - DB^* - [c,\psi ^*]                     \cr
    s\phi ^* &= - D\psi ^* - {1\over 2}[B^*,B^*] - [c,\phi ^*]\ ,\cr
 }\right.
 \eqno(2.13{\rm a})
$$
$$
  \tilde D\tilde B=0\quad \Rightarrow \
\left\{ \eqalign
 {
    sc^* &=  - DA^*  - [B^*,B] - [\psi ^*,\psi ]
               - [\phi ^*,\phi ] - [c,c^*] \cr
    sA^* &=  - DB   - [B^*,\psi ] - [\psi ^*,\phi ] - [c,A^*]     \cr
    sB   &=  - D\psi - [B^*,\phi ] - [c,B]                        \cr
    s\psi  &=  - D\phi - [c,\psi ]                                \cr
    s\phi  &=  - [c,\phi ] \quad ,                                \cr
 }\right.
 \eqno(2.13{\rm b})
$$
which is a total BRST algebra involving the BV antifields
and is off-shell nilpotent.

It is a distinctive advantage in our method that
a BRST invariant BV action is obtained quit easily as follows.
We present a BV antifield action
$$
 {\cal S}_{\rm BV}= \int _{M_4}{\rm Tr} \left( \tilde B \wedge \tilde F
                                    - \tilde B \wedge s\tilde A \right)
 \eqno(2.14)
$$
by means of the extended forms, which turns out to be equivalent to
$$
 \eqalign
 {
 {\cal S}_{\rm BV}=
 \int _{M_4} \Tr
  \biggl[
   (B \wedge F ) &+ A^* \wedge Dc + c^* \wedge {1\over 2}[c,c] \cr
         &+ B\wedge [c,B^*] + \psi \wedge ( DB^* + [c,\psi ^*] ) \cr
         &+ \phi \wedge ( D\psi ^*  + {1\over 2}[B^*,B^*] + [c,\phi ^*] )
  \ \biggr] \cr
 }
 \eqno(2.15)
$$
in its component fields.
It is obvious that the BRST transformations (2.13a,b) are produced
by this action ${\cal S}$ as a generator
according to the BV antifield-antibracket formalism, that is,
$\ s\Phi = (\Phi ,{\cal S}) = {\delta {\cal S}/\delta \Phi ^*} \ $,
$\ s\Phi ^* = (\Phi ^*,{\cal S}) = {\delta {\cal S}/\delta \Phi } \ $.
Hence it follows that the action (2.15) is nothing but a proper
solution for the master equation with its minimal constituent.
Though a superficial reversal of signs may occur,
it is due to a difference of conventions, that is to say,
the BRST operator in the BV formalism
is defined as a operation
from the right, while our operation is from the left.

\vskip20pt
\noindent{\bf 3. Odd canonical formalism and BV action}

Let us inquire into the meaning of our BV action (2.14) and the relation to
the BRST conditions (2.12a,b).
One may consider an action
$$
  \tilde S[\tilde A,\tilde B]
 = \int _{M_4}\Tr\left( \tilde B \wedge \tilde F \right)
 \eqno(3.1)
$$
which is extended from the invariant $BF$ action (2.1) by
substituting the extended forms $\tilde A $ and $\tilde B $ for
$A$ and $B$.
It is obvious that the conditions (2.12a,b)
deriving the BRST algebra are nothing but formal field equations
from this extended action.
Though the extended action itself is not our object,
our BV action seems to be something like a Legendre transform of it.
Here we imply the Legendre transform of
$\ s\tilde A \ $ to $\ {\tilde A}^* = \tilde B$,
which is an analogy of translation from a Lagrangian to a Hamiltonian
provided that $ s\tilde A $ pretends to be a time derivative of $\tilde A$,
though it is an odd time.
To say things more clear,
if we define odd ``canonical momenta'' $\tilde \pi _A$ and $\tilde \pi _B$
of $\tilde A$ and $\tilde B$ as
$$
 \tilde \pi _A
 := {\partial \ \tilde {\cal L}\over \partial (s\tilde A)}
  = \tilde B \ ,\quad
 \tilde \pi _B
 := {\partial \ \tilde {\cal L}\over \partial (s\tilde B)} = 0   \ ,
 \eqno(3.2{\rm a,b})
$$
provided $\  \tilde S = \displaystyle\int _{M_4}\tilde {\cal L}\ $,
then it follows that ``Hamiltonian'' $\tilde {\cal H}$ is defined by
$$
  \tilde {\cal H}
 := \Tr\left( \tilde \pi _A \wedge s\tilde A
      + \tilde \pi _B \wedge s\tilde B \right) - \tilde {\cal L}
  = \Tr\left( \tilde B\wedge s\tilde A
    - \tilde B \wedge \tilde F \right) \ .
 \eqno(3.3)
$$
The BV action presented in the previous section is
nothing but this ``Hamiltonian'',
provided that its total sign is changed
and it is integrated on the manifold, that is,
$$
 {\cal S}_{\rm BV}= - \displaystyle\int _{M_4}\tilde {\cal H}\ .
 \eqno(3.4)
$$

 The odd time canonical formulation of the BV formalism can be found
in ref. [13].
The BV antifield $\Phi ^*$ plays a role of canonical momentum
of the field $\Phi $ and the antibracket acts as an odd Poisson bracket
in this odd canonical formalism. At this time, the BV action
${\cal S}(\Phi ,\Phi ^*)$ takes the place of a Hamiltonian and
the BRST transformation of fields is
just an odd time evolution equation brought by this ``Hamiltonian''.
The BV master equation $({\cal S},{\cal S})=0$
means a constancy of the ``Hamiltonian'' ${\cal S}$ through the
``time evolution''.

Our method may clarify the major unsatisfactory point in the odd time
formalism.
They explain in the ref.[13] that
it is not clear if the odd time formulation can
give some indication for solving the master equation, and beside
there is hardly any concrete discussion about
the ``Lagrangian'' and its role
except for the assumption that it exists.
One may see that the point which have been missed is the  extended
Lagrangian and that the ``Hamiltonian'' obtained from it is nothing but
a BV action that is provided as a solution for the master equation.
There seems to be a close connection between
our method and the odd time formulation,
which we will discuss in detail elsewhere.

\vskip20pt
\noindent{\bf 4. $BF$ theory in arbitrary dimensions}

 It is straightforward to generalize our construction
for the $BF$ theory to the model in arbitrary dimensions.
The invariant action  for the $BF$ theory in $D$-dimensional
spacetime has the same form as the previous one,
$$
  S = \int _{M_D}{\rm Tr} ( B_n \wedge F )
 \eqno(4.1)
$$
provided that $B_n$ is a $n$-form in this case,
where we assume $\ n=D-2,\ D\geq 4 $.
There is a symmetry with $(n-1)$-form $\varepsilon _{n-1}$
$$
  \delta _{\varepsilon _{n-1}}A
= 0\ ,\quad \delta _{\varepsilon _{n-1}}B_n = D\varepsilon _{n-1}\ ,
 \eqno(4.2)
$$
which is reducible when $\varepsilon _{n-1}=D\lambda _{n-2}$\ .
This reducible symmetry requires a sequence of ghosts descended from
$\ B_n=B_{(n,0)}\ $:
$$
  B_{(n,0)}\rightarrow B_{(n-1,1)}\rightarrow
\ldots \rightarrow B_{(n-q,q)}\rightarrow \ldots \rightarrow B_{(0,n)}\ ,
\eqno(4.3)
$$
and there is also an required  ordinary ghost $c$ for Yang-Mills symmetry.
The extended forms should be defined as
$$
 \eqalignno
 {
  \tilde A   &= c_{(0,1)} + A_{(1,0)}
            + \sum _{q=0}^n B^*_{(2+q,-1-q)}    \ ,
        &(4.4{\rm a}) \cr
  \tilde B_n &= c^*_{(n+2,-2)}
   + A^*_{(n+1,-1)} + \sum _{q=0}^n B_{(n-q,q)}\ ,
        &(4.4{\rm b}) \cr
 }
$$
including BV antifields,
hence they can be regarded as antifield conjugates of each other.
The conditions for BRST transformations are again
$$
  \tilde F=0\ ,\quad \tilde D\tilde B_n=0\ ,
 \eqno(4.5{\rm a,b})
$$
and the  BV action is of the same form as the previous one
$$
 {\cal S}_{\rm BV}= \int _{M_D}{\rm Tr}
 ( \tilde B_n\wedge \tilde F - \tilde B_n \wedge s\tilde A)\
 \eqno(4.6)
$$
in terms of the extended forms.
Expansion of the conditions (4.5a,b)
in the ghost number derives the BRST transformations
$$
 \left\{
 \eqalign
  {
    sc  \ \quad &= - {1\over 2}[c,c]
   \cr
    sA  \quad \quad &= - Dc
   \cr
    sB^*_{(2,-1)} \ &= - F - [ c, B^*_{(2,-1)} ]
   \cr
    sB^*_{(3,-2)} \quad &= - DB^*_{(2,-1)} - [ c, B^*_{(3,-2)} ]
   \cr
    sB^*_{(2+q,-q-1)}
   &=  - DB^*_{(1+q,-q)}
    -{1\over 2}\sum _{q'=0}^{q-2}
   [ B^*_{(2+q',-1-q')}, B^*_{(q-q',q'+1-q)}]
   \cr
   &\hskip120pt
    - [ c, B^*_{(2+q,-1-q)} ]\ ,\quad (2\leq q\leq n)\ ,
   \cr
 }\right.
 \eqno(4.7{\rm a})
$$
$$
 \left\{
 \eqalign
 {
   sc^* \quad
  &=   - DA^* - \sum _{q'=0}^{n} [ B^*_{(2+q',-1-q')}, B_{(n-q',q')} ]
           - [c, c^*]
  \cr
   sA^* \quad
   &=  - DB_{(n,0)}
       - \sum _{q'=0}^{n-1} [ B^*_{(2+q',-1-q')}, B_{(n-1-q',1+q')} ]
       - [c,A^*]
  \cr
   sB_{(n-q,q)}  &=  - DB_{(n-q-1,q+1)}
    - \sum _{q'=0}^{n-q-2} [ B^*_{(2+q',-1-q')}, B_{(n-q-2-q',q+2+q')} ]
   \cr
    &\hskip150pt - [c,B_{(n-q,q)} ] \ ,\quad ( 0\leq q\leq n-2)
  \cr
   sB_{(\ 1,n-1)} &=  - DB_{(0,n)} - [c,B_{(1,n-1)} ]
  \cr
   sB_{(\ 0,n)} \quad &=  -[c,B_{(0,n)} ]\quad .
  \cr
 }\right.
 \eqno(4.7{\rm b})
$$
The BRST invariant BV action that involves the antifields
is obtained by the explicit calculation of the action (4.6),
$$
 \eqalign
 {
 {\cal S}_{\rm BV}=
\int _{M_D}\biggl[
  &\left( B_{(n,0)}\wedge F_{(2,0)}\right)
         + c^* \wedge {1\over 2}[c,c] + A^* \wedge Dc
  \cr
 &+ B_{(n,0)} \wedge [ c, B^*_{(2,-1)} ]
  + B_{(n-1,1)}\wedge \left( DB^*_{(2,-1)} + [ c, B^*_{(3,-2)} ]   \right)
  \cr
  &+ \sum _{q=2}^n B_{(n-q,q)}\wedge
  \cr
    \biggl(
    DB^*_{(1+q,-q)}
   &
      +{1\over 2}\sum _{q'=0}^{q-2}
         [ B^*_{(2+q',-1-q')}, B^*_{(q-q',q'+1-q)}]
    + [ c, B^*_{(2+q,-1-q)} ]
     \biggr)
  \biggr]\ .
  \cr
 }
 \eqno(4.8)
$$

 The action obtained above is just a minimal solution of
the BV master equation, hence there should be introduced
the antighosts and multipliers in order to fix the symmetries.
Then the gauge fixed BRST invariant action will be obtained by
the substitution of the antifields with
$\ \Phi ^* = \displaystyle{\partial \Psi \over \partial \Phi }\ $
taking a suitable gauge fermion $\Psi $.
It happens at this stage that the BRST transformation may be nilpotent
only on-shell, when the BV action contains
quadratic or higher order terms in the antifields,
because the BRST transformation after gauge fixing is defined by
$$
  \delta _{\rm BRST}\Phi
 = \left( \Phi ,{\cal S}\right) \bigg|_{\Phi ^*
 = {\partial \Psi \over \partial \Phi }}
 = {\partial {\cal S}\over \partial \Phi ^*}
 \bigg|_{\Phi ^* = {\partial \Psi \over \partial \Phi }}\ .
 \eqno(4.9)
$$

 It appears that the models we are dealing up to now, namely,
the non-abelian $BF$ theories in 4 and higher dimensions
suffer from such a defect.
The BRST algebra after gauge fixing is on-shell nilpotent,
because the BV action (4.8)
contains quadratic terms in the antifields $B^*$.
As a consequence, the gauge fixed BRST operator
$\ \delta _{\rm BRST} \ $
is on-shell nilpotent and requires the $B$ field equations
to be satisfied
when it acts on them.
Although one may think that the elimination of $B$ fields
by substituting their equations possibly cures the situation,
it seems in fact difficult to accomplish such a procedure in this case.

Though it is of course a significant issue to be settled,
we will not pursue the problem any more in this paper,
because it is a little way off our main theme.

\vskip20pt
\noindent{\bf 5. Chern-Simons theory}

 Our method of extended forms can be applicable to
some other theories as far as TQFT are concerned.
Let us consider how it works in case of the Chern-Simons theory in
3-dimensions whose action is the form of
$$
  S[A] = {1\over 2}\int _{M_3}{\rm Tr}
\left( A\wedge dA + {2\over 3}A\wedge A\wedge A \right) \ .
 \eqno(5.1)
$$
It is required only to introduce an ordinary ghost $c$, because
the only symmetry of the action, that is,
an ordinary Yang-Mills gauge symmetry is not reducible.
We define an extended 1-form $\tilde A$ by
$$
  \tilde A = c_{(0,1)} + A_{(1,0)} + A^*_{(2,-1)} + c^*_{(3,-2)}
 \eqno(5.2)
$$
including BV antifields $A^*$ and $c^*$ as well as $A$ and $c$.
The extended action $\tilde S[\tilde A] $ is formally obtained by replacing
$A$ with $\tilde A$ together with $\tilde d$ substituted for $d$
in the Chern-Simons action,
$$
  \tilde S[\tilde A] = {1\over 2}\int _{M_3}{\rm Tr}
                \left( \tilde A\wedge \tilde d\tilde A
  + {2\over 3}\tilde A\wedge \tilde A\wedge \tilde A \right) \ ,
 \eqno(5.3)
$$
whose formal field equation
$$
  \tilde F=0
 \eqno(5.4)
$$
derives the BRST transformations
$$
\left\{
 \eqalign
 {
   sc   &= -{1\over 2}[c,c]      \cr
   sA   &= -Dc               \cr
   sA^* &= - F - [c,A^*]     \cr
   sc^* &= - DA^* - [c,c^*]  \ ,\cr
 }
 \right.
 \eqno(5.5)
$$
when it is expanded in ghost number.

 According to the procedure described in \S3,
we define an odd ``momentum''$\tilde \pi $ of $\tilde A$ by
$$
 \tilde \pi := {\partial \tilde {\cal L}\over \partial (s\tilde A)}
= {1\over 2}\tilde A
 \eqno(5.6)
$$
and a ``Hamiltonian'' $\tilde {\cal H}$ by
$$
 \tilde {\cal H} := \Tr(\tilde \pi \wedge s\tilde A) - \tilde {\cal L}\ ,
 \eqno(5.7)
$$
provided $\ \tilde S[\tilde A]= \displaystyle\int _{M_3}\tilde {\cal L}\ $.

The BRST invariant BV action is nothing but this ``Hamiltonian'',
provided that its sign is changed and it is integrated on the manifold,
$$
 {\cal S}_{\rm BV}
 = - \int _{M_3}\tilde {\cal H}
 \ =  \int _{M_3} {\rm Tr}
  \left[ {1\over 2}
\left( \tilde A\wedge \tilde d\tilde A
+ {2\over 3}\tilde A\wedge \tilde A \wedge \tilde A \right)
- {1\over 2}\tilde A \wedge s\tilde A \right] \ .
 \eqno(5.8)
$$
The explicit calculation in its components
would show us that the action has the form of
$$
 {\cal S}_{\rm BV}
 =  \int _{M_3} {\rm Tr}
  \left[ {1\over 2}\left(
  A\wedge dA + {2\over 3}A\wedge A\wedge A \right)
 + A^* \wedge Dc + c^* \wedge {1\over 2}[c,c] \right] \ .
 \eqno(5.9)
$$

%\vskip20pt
\endpage

\noindent{\bf 6. $BF$ theory with ``cosmological term'' and
 topological Yang-Mills theory}

 There is considered a kind of extension
of the 4-dimensional $BF$ theory to include a term similar
to the cosmological term [9], whose action is given by
$$
  S[A,B] = \int _{M_4}{\rm Tr}
\left( B \wedge F - {\Lambda \over 2}B\wedge B\right)
 \eqno(6.1)
$$
with a constant $\Lambda $ resembling  the cosmological constant.

The theory described by the above action
(which we will call $BF_\Lambda $ theory)
is also topological and has a symmetry
$$
  \delta _\varepsilon A = \Lambda \varepsilon \ ,
   \quad \delta _\varepsilon B = D\varepsilon \ ,
 \eqno(6.2)
$$
besides the ordinary Yang-Mills symmetry.
The symmetry (6.2) allows us to transform the connection $A$
to an arbitrary form and makes the theory topological.
As a matter of fact, the $BF_\Lambda $ action (6.1) is equivalent
to the underlying action of topological Yang-Mills (TYM) theory,
$$
  S[A] = {1\over 2\Lambda }\int _{M_4}{\rm Tr} \left( F\wedge F\right) \ ,
 \eqno(6.3)
$$
when we substitute a field equation for $B$.
The field equations of the action (6.1) are
$$
  F - \Lambda B = 0\ ,\quad DB=0\ ,
 \eqno(6.4)
$$
the latter of which is just a consistency of
the former with the Bianchi identity $\ DF=0 \ $.
It is well known that a BRST invariant gauge fixed action
starting from (6.3) describes the TYM theory.

 In order to construct a BRST invariant BV action,
we define the extended forms
$$
\eqalignno
 {
 \tilde A &= c_{(0,1)} + A_{(1,0)}
       + B^*_{(2,-1)} + \psi ^*_{(3,-2)} +\phi ^*_{(4,-3)}
     &(6.5{\rm a})\cr
 \tilde B &= c^*_{(4,-2)} + A^*_{(3,-1)}
         + B_{(2,0)} + \psi _{(1,1)} +\phi _{(0,2)}
     &(6.5{\rm b})\cr
 }
$$
and the extended action
$$
  \tilde S[\tilde A,\tilde B]
= \int _{M_4}{\rm Tr}\left( \tilde B\wedge \tilde F
              - {\Lambda \over 2}\tilde B\wedge \tilde B\right)
 \eqno(6.6)
$$
in the same way as the $BF$ theory in \S2.
The formal field equations
$$
  \tilde F - \Lambda \tilde B = 0\ ,\quad \tilde D\tilde B=0\ ,
 \eqno(6.7{\rm a,b})
$$
of this action are conditions that produces BRST algebra as follows.
$$
 \kern-90pt
 \tilde F - \Lambda \tilde B =0\quad \Rightarrow \
 \left\{ \eqalign
  {
    sc    &= \Lambda \phi - {1\over 2}[c,c]                      \cr
    sA    &= \Lambda \psi - Dc                                   \cr
    sB^* &= - ( F - \Lambda B)- [c,B^*]                         \cr
    s\psi ^* &= \Lambda A^* - DB^* - [c,\psi ^*]                \cr
    s\phi ^* &= \Lambda c^* - D\psi ^*  - {1\over 2}[B^*,B^*]
                                                - [c,\phi ^*]\ ,\cr
  } \right.
 \eqno(6.8{\rm a})
$$
$$
 \tilde D\tilde B = 0 \quad \Rightarrow \
 \left\{ \eqalign
  {
   sc^* &= - DA^*  - [B^*,B] - [\psi ^*,\psi ]
                  - [\phi ^*,\phi ] - [c,c^*] \cr
   sA^* &= - DB   - [B^*,\psi ] - [\psi ^*,\phi ] - [c,A^*] \cr
   sB  &= - D\psi   - [B^*,\phi ] - [c,B]  \cr
   s\psi  &= - D\phi  - [c,\psi ]    \cr
   s\phi  &=  - [c,\phi ]          \quad .\cr
  } \right.
 \eqno(6.8{\rm b})
$$
It is easy to see that the BRST above does not differ from
that of topological Yang-Mills theory,
when we eliminate $B$ and $B^*$ provided $\Lambda =1$.

The BV action with antifields is obtained in the same way as before.
We define the odd ``canonical momenta $\tilde \pi _A$ and $\tilde \pi _B$
for $\tilde A$ and $\tilde B$ by
$$
 \tilde \pi _A
 := {\partial \tilde {\cal L}\over \partial (s\tilde A)}
 = \tilde B \ ,\quad
 \tilde \pi _B
 := {\partial \tilde {\cal L}\over \partial (s\tilde B)} = 0   \ ,
 \eqno(6.9{\rm a,b})
$$
and define the ``Hamiltonian'' $\tilde {\cal H}$ by
$$
  \tilde {\cal H}
 := \Tr\left( \tilde \pi _A \wedge s\tilde A
        + \tilde \pi _B\wedge s\tilde B\right) - \tilde {\cal L}
     \ = \Tr\left[ \tilde B \wedge s\tilde A
 - \left( \tilde B\wedge \tilde F - {\Lambda \over 2}
             \tilde B\wedge \tilde B\right) \right] \ ,
 \eqno(6.10)
$$
then the BV action is given by
$$
 {\cal S}_{\rm BV}
   = - \int _{M_4}\tilde {\cal H}
 \ =  \int _{M_4} \Tr\left[ \left( \tilde B\wedge \tilde F
             - {\Lambda \over 2}\tilde B\wedge \tilde B\right)
                             -\tilde B \wedge s\tilde A \right]
 \eqno(6.11)
$$
in the extended forms, which turns out to be
$$
 \eqalign
 {
  {\cal S}_{\rm BV}=
  \int _{M_4} \Tr
  \biggl[
  \left( B\wedge F - {\Lambda \over 2}B\wedge B\right)
    + A^* \wedge ( \Lambda \psi + Dc ) + c^* \wedge
               \left( \Lambda \phi + {1\over 2}[c,c] \right)
   \quad \cr
    + B \wedge [c,B^*] + \psi \wedge \left( DB^* + [c,\psi ^*] \right)
  \quad \cr
   + \phi \wedge
 \left( D\psi ^*  + {1\over 2}[B^*,B^*] + [c,\phi ^*] \right)
  \biggr]
  \cr
 }
 \eqno(6.12)
$$
by expanding in the ghost number.

 This is a minimal solution of the BV master equation,
hence the BRST invariant gauge-fixed action is obtained
after introducing the antighosts and multipliers,
and the gauge fermion is also required as explained in \S5.
It seems that the BRST algebra becomes on-shell nilpotent at this stage.
The gauge fixed BRST operator $\ \delta _{\rm BRST}\ $
is nilpotent on $B$ equation,
because there is a quadratic term of $B^*$ in the BV action (6.12).

The elimination of the $B$ field by substituting its equation of motion
may not cause any trouble in this case
on the contrary to the previous case of $BF$ theory.
For example, one may take a gauge condition which does not
contain $B$ field for simplicity
such as $\ {}^- F=0\ ,\ \partial \!\cdot \!A=0\ $
employed in the topological Yang-Mills theory.
It requires $B^*=0$ and allows us to eliminate $B$
by substitution of its equation
$\ F - \Lambda B=0\ $, then we obtain a gauge fixed action
which is equivalent
to that of the TYM theory, provided $\Lambda =1$.
In this way, the requirement for the off-shell nilpotent BRST algebra
in the $BF_{\Lambda }$ theory
renders the theory equivalent to the TYM theory.
As a matter of fact, if we start from the action
$$
  S[A] = {1\over 2}\int _{M_4}{\rm Tr} \left( F\wedge F\right)
 \eqno(6.14)
$$
and apply our extended form method to it, then the resulting BV action
is the same form as that of the $BF_{\Lambda }$
theory  (in case of $\Lambda =1$) provided
that the odd momentum
$\tilde \pi _A$ does actually have the same form
as $\tilde B$ in its components.

Though the $BF_{\Lambda }$ theory
is almost equivalent to the TYM theory when $\Lambda \not=0$,
in the $\Lambda \rightarrow 0$ limit it becomes the usual $BF$ theory
which is akin to the Chern-Simons theory.
It is said that the TQFT are classified into two types of theories,
one of which is the Witten type TQFT ({\it e.g.} TYM theory)
and the other is the Schwartz type TQFT
({\it e.g.} Chern-Simons theory or $BF$ theory) [4].
The Witten type is often called ``cohomological'',
whose classical action is a topological index or simply zero and
quantum action is just a BRST commutator.
The Schwartz type is often called ``quantum'', whose classical action
is nontrivial and a quantum action is not merely a BRST commutator.
It is interesting that the $BF_{\Lambda }$ theory may possibly incorporate
the different kinds of TQFT as limiting cases.
We will discuss these features in detail elsewhere.

\vskip20pt
\noindent{\bf 7. Conclusion}

 We have proposed an extended differential calculus to construct
BV antifield actions for the TQFT.
Though our method is constructed on the basis similar to
the universal bundle method in the TYM theory,
it is rather extended so as to include
all the required ghosts and the BV antifields.
The extended $r$-form is defined by
$$
  \tilde \Phi _r = \sum _{r=p+q} \Phi _{(p,q)}\ ,
 \eqno(7.1)
$$
where the sum is taken over the forms of degree $p=0$ to $D$,
(that is, the dimensionality of the manifold),
which allows the ghost number $q$
to be negative and consequently the BV antifields are
included as well as the ghosts.
It is quite easy to obtain
not only the conditions that derive the BRST algebra
but also the BRST invariant BV action itself
owing to the fully extended forms.

 The method we have proposed in this paper may be summarized into the
steps as follows,
which is something like a procedure in the canonical formalism,
provided that  $\ s\ $ pretends to behave an odd time derivative.
The first step is to obtain an extended action by
substituting the forms with the extended ones:
$$
 S[\Phi ] = \int _{M}{\cal L}(\Phi ,d\Phi )
 \quad \rightarrow \quad \tilde S[\tilde \Phi ]
=\int _{M}\tilde {\cal L}(\tilde \Phi ,\tilde d\tilde \Phi ) \ .
 \eqno(7.2)
$$
The second step is  the ``Legendre transform''
from the extended Lagrangian to the ``Hamiltonian''
by defining the odd ``canonical momentum'':
$$
 \eqalign
 {
 \tilde {\cal L}(\tilde \Phi , d\tilde \Phi , s\tilde \Phi )
 \ \rightarrow \
 \tilde {\cal H}(\tilde \Phi , d\tilde \Phi ,\tilde \pi )
 & := \tilde \pi \wedge s\tilde \Phi - \tilde {\cal L} \ ,
 \cr
 &\left( \tilde \pi :=
 {\partial \ \tilde {\cal L}\over \partial (s\tilde \Phi )}\right) \quad .
 \cr
 }
 \eqno(7.3)
$$
At the final step,
we change the sign of this ``Hamiltonian''
and integrate it on the manifold,
which is nothing but the BV antifield action that is a minimal
solution for the master equation:
$$
 {\cal S}_{\rm BV}[\Phi ,\Phi ^*]
= -\int _{M}\tilde {\cal H}(\tilde \Phi , d\tilde \Phi ,\tilde \pi )
 \eqno(7.4)
$$

The conditions that derive the BRST algebra
are the equation of motion from the extended action,
which is equivalent to
the odd time evolution equations brought by this ``Hamiltonian''.

Although it is not yet clear whether it is applicable to
the non-topological theories or not,
it seems that our method can be applied to other TQFT as far as they are
written down by means of the differential forms
without referring the metric.
A detailed investigation on general structure of these method will be
necessary.

\vskip20pt
\noindent{\fourteenbf Acknowledgement}

The author would like to thank S. Tsujimaru for fruitful discussions.
It is also a pleasure to thank S. Kitakado for reading
the manuscript and for encouragement.

%\endpage
\vskip40pt

\centerline{\seventeenbf References}

\item{[1]}
E. Witten, Commun. Math. Phys. 117 (1988) 353.

\item{[2]}
J.M.F. Labastida and M. Pernici, Phys. Lett. B 212 (1988) 56

\item{[3]}
D. Birmingham, M. Rakowski and G. Thompson, Nucl. Phys. B315 (1989) 577.

\item{[4]}
D. Birmingham, M. Blau, M. Rakowski and G. Thompson,
Phys. Rep. 209 (1991) 129, and references therein.

\item{[5]}
I.A. Batalin and G.A. Vilkovisky, Phys. Lett. B 102 (1981) 27,  \hfill\break
I.A. Batalin and G.A. Vilkovisky, Phys. Rev. D 28 (1983) 2567,  \hfill\break
I.A. Batalin and G.A. Vilkovisky, J. Math. Phys. 26 (1985) 172.

\item{[6]}
H. Kanno, Z. Phys. C 43 (1989) 477.

\item{[7]}
L. Baulieu and I.M. Singer, Nucl. Phys. (Proc. Suppl.) 5B (1988) 12.

\item{[8]}
S. Ouvry, R. Stora and P. van Baal, Phys. Lett. B 220 (1989) 159.

\item{[9]}
G. Horowitz, Commun. Math. Phys. 125 (1989) 417.

\item{[10]}
M. Blau and G. Thompson, Phys. Lett. B 228 (1989) 64.

\item{[11]}
J.C. Wallet, Phys. Lett. B 235 (1990) 71.

\item{[12]}
H. Ikemori, Shiga University preprint, Shiga-92-1.

\item{[13]}
{\"O}.F. Dayi, Mod. Phys. Lett. A 4 (1989) 361.

\end